\documentclass[]{tMOP2e}

\usepackage{multicol}
\usepackage{multirow}
\usepackage{booktabs}
\usepackage{graphicx}
\usepackage{epstopdf}
\usepackage{subfigure}

\theoremstyle{plain}

\theoremstyle{remark}

\begin{document}


\title{Radiation tolerance of opto-electronic components proposed for space-based quantum key distribution}

\author{Yue Chuan Tan$^{\ast}$\thanks{$^\ast$Corresponding author. Email: cqttyc@nus.edu.sg
\vspace{6pt}}, Rakhitha Chandrasekara,  Cliff Cheng and Alexander Ling\\\vspace{6pt}  
$^{a}${\em{Centre for Quantum Technologies, National University of Singapore, Block S15, 3 Science Drive 2, Singapore 117543}}} 


\maketitle

\begin{abstract}
Plasma in low earth orbit can damage electronic components and potentially jeopardise scientific missions in space.
Predicting the accumulated damage and understanding components' radiation tolerance are important to mission planning. 
In this manuscript we report on the observed radiation tolerance of single photon detectors and a liquid crystal polarization rotator.
We conclude that an uncooled Si APD could continue to operate from more than a month up to beyond the lifetime of the satellite depending on the orbit. The liquid crystal polarization rotator was also unaffected by the exposed dosage.


\end{abstract}

\section{Introduction}
\label{sec:intro}
The requirement for line-of-sight transceiver locations limits the range of terrestrial free-space optical quantum key distribution (QKD) \cite{ursin07} .
This limitation may be overcome, hence establishing a global QKD network by using satellites \cite{Scheid2013} in Low Earth Orbit (LEO) to transmit quantum states of light through an optical link.
For discrete variable QKD employing polarization entangled photons, single photon sources and detectors can be carried on a satellite.
The first step in this direction is to demonstrate that the necessary opto-electronics components in a space-based QKD network will operate in LEO.
We have proposed that this can be performed cost-effectively with nanosatellites called cubesats \cite{morong12}.
Cubesats in LEO (at an approximate altitude of 400\,km) will be able to survive for one year in orbit before falling back to Earth. 
We anticipate that one year is sufficient time to carry out a proper space-based QKD demonstration.

The basic unit of a cubesat is a 10\,cm cube (1U) with mass below one kilogram and with typically 1.5\,W of electrical power available \cite{woellert11}. Each 1U cube may be stacked into larger satellites and typically are designed around commercial-off-the-shelf technologies. However, as the size and mass available are limited, there is little room for adding shields against space radiation. It is necessary to ensure that the opto-electronics devices in the experiment can operate reliably for the expected lifetime of one year in LEO. 


Space radiation in LEO consists of mainly protons and electrons trapped in the Earth's magnetic fields \cite{sun97}.
Proton induced damage is primarily displacement damage, while electrons tend to cause ionizing damage.
Both damage mechanisms result in degradation and operational failure of spacecraft electronics. 
For a space-based QKD experiment, the most radiation sensitive opto-electronic devices are Si avalanche photodiodes (APDs) for detecting single photons and liquid crystal-based polarization rotators (LCPR) used for selecting the polarization basis of the QKD measurements.

The target orbit for our proposed missions is at an altitude of about 400\,km.
More launch opportunities are available at this altitude, including possible deployment via the International Space Station (ISS). 
Furthermore, satellites at this altitude fall back to Earth within a year and so there is no risk of long-term space debris from our proposed mission.
To test the radiation tolerance of the components, the target radiation environment was modeled using the Space Environment Information System (SPENVIS) \cite{spenvis} with 2\,mm of aluminium shielding assumed.
The simulation results from SPENVIS were used to determine the levels of irradiation for the test components.
In this paper, we present data on the performance of the APDs after exposure to proton irradiation as well as data on the performance of the LCPR after exposure to ionizing radiation.

\section{Avalanche Photodiodes}
\label{sec:APD}

\begin{figure}
\begin{center}
\includegraphics[width=\textwidth]{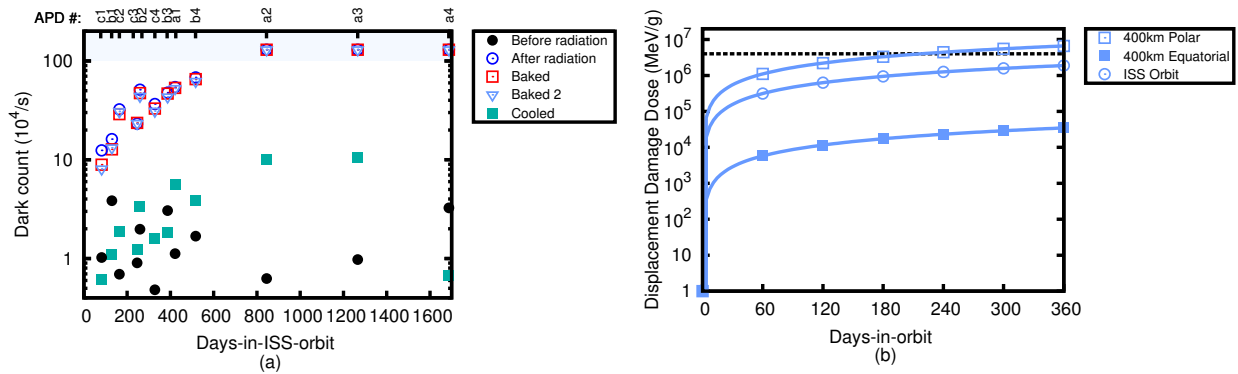}
\caption{(a) Dark count rates pre- and post-irradiation of twelve APDs. 
Each column of points corresponds to the dark counts of one APD that had been subjected to a particular proton fluences (converted into days in ISS orbit, see Fig. \ref{fig:APDcombined}(b)) at various experiment stages.
After irradiation, devices were annealed at 55\,$^\circ$C for 24 hours revealing a small but permanent drop in the dark count rate. The biggest drop in dark count rates was observed when the devices were cooled to -20\,$^\circ$C. (b) Expected Proton Displacement Damage Dose (DDD) for three different orbits. Solid blue lines are linear fit for the DDD in various orbits. These plots reveal that there is a strong dependence on the orbit inclination with the lowest radiation fluences experienced in an equatorial orbit. The black dashed line is the approximate guide where the accumulated displacement damage starts to saturate the detectors operating at room temperature (based on the test result of APD\,\#: a2 in Fig. \ref{fig:APDcombined}(a)).}
\label{fig:APDcombined}
\end{center}
\end{figure}

Twelve Silicon APD devices (SAP500, 500\,$\mu$m active area) were exposed to different levels of proton irradiation at the Crocker Nuclear Laboratory (CNL) in the University of California at Davis. 
The displacement damage dose (DDD) is a measurement of the total accumulated damage, and is a function of both the proton energy and the total fluence, given by
\begin{eqnarray}
\mathrm{DDD}= \phi_\mathrm{p}(E)S_\mathrm{p}(E)
\label{eqn:ddd}
\end{eqnarray}
where $\phi_\mathrm{p}$ is the fluence of protons with energy E, and $S_\mathrm{p}$ is the corresponding NIEL \cite{Messenger2001}.
A detailed discussion on the estimation of the DDD and its conversion to equivalent days in orbit is available at \cite{Tan2013}. 

The dark count rate before and after irradiation is shown in Fig.\,\ref{fig:APDcombined}(a).
The dark counts of uncooled (22-25\,$^\circ$C) APDs are expected to saturate the passive quenching circuit after a dose of 5$\times 10^{6}$\,MeV/g, corresponding to approximately 700 days in ISS orbit. 
Annealing the devices at 55\,$^\circ$C for 24 hours has a small but permanent recovery effect.
However, beyond a DDD of 3$\times 10^{6}$\,MeV/g, annealing appears to be ineffective. 
Cooling the devices (-20\,$^\circ$C) reduces dark counts by 2 orders of magnitude, suggesting that long-running space-based experiments employing APDs should investigate the use of devices which come with integrated thermoelectric coolers. 
Based on the measurement results, the acceptance level for proton damage of the uncooled SAP500 APDs is found to be 5$\times 10^{6}$\,MeV/g of DDD. 

Earlier studies also reported similar results where the proton radiation causes an increase in APD dark count rates and the rate could be partially reduced by thermal annealing \cite{Kodet2012,Moscatelli2013}. However, we note that thermal annealing appears to be ineffective after the APDs were subjected to a DDD of 3$\times 10^{6}$\,MeV/g, which is an observation that has not been previously reported. This may have been missed in earlier reports due to the sub-zero operating temperature \cite{Moscatelli2013} or the use of small active area devices \cite{Kodet2012}.

In contrast to the deep space missions where APD cooling is necessary to prolong its effective lifetime in orbit, our cubesat missions are of much shorter lifetime, typically on the order of a year. Therefore, we are able to get away with APD cooling. It is also worth pointing out that we use free running passive quenched circuits to drive the APD while APDs in the other two studies were actively quenched and gated. This could also explain the reason their APDs remains operational at higher dark count rates (on the order of Mcps) as reported in \cite{Kodet2012} while our APDs saturates the circuit and became unusable beyond dark counts rates of 1Mcps. The disadvantage of active quench and sub-zero temperature operation is that the power budget exceeds the capability of standard cubesats. Our selected APDs should also reach the performance level of the devices in \cite{Kodet2012,Moscatelli2013} if we had access to spacecraft that can provide a larger energy budget.

The rate of displacement damage dose accumulation in different orbits is compared in Fig.\,\ref{fig:APDcombined}(b). Polar orbits and equatorial orbits are defined as orbits having inclinations of 98 and 20 degrees, while the ISS orbit, which is the orbit of greatest interest, has an inclination of 45 degrees. 
It is evident that inclination of the orbit is a major factor on the radiation environment with the polar orbits being harsher than equatorial orbits.
The analysis suggests that long duration photon counting experiments can be conducted at 400 km with low inclination orbits that even uncooled APDs can operate beyond the lifetime of the spacecraft. Quantum experiments relying on single photon counting may be operated at all orbits if sufficient power is supplied for cooling using integrated thermoelectric coolers. This approach may also enable annealing cycles for recovery from proton-induced damage.  Finally, devices with smaller active areas may be investigated, as the reduced cross-section is expected to accumulate less damage for a given fluence.

\section{Liquid Crystal Polarization Rotator}
\label{sec:LCD}

\begin{figure}
\begin{center}
\includegraphics[width=0.9\textwidth]{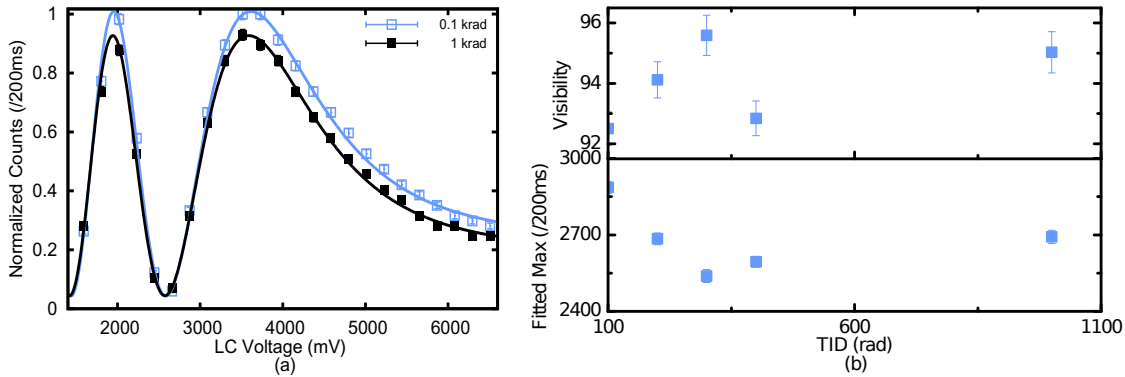}
\caption{(a) Normalized detected single photons after correcting for the background signal at different applied voltage to the liquid crystal-based polarization rotator (LCPR). The data is fitted with a retarded sinusoid, the anticipated polarization rotation behaviour of the LCPR devices.  (b) Visibility of each set of measurements after correction was determined. The signal strength is given by the maximum of the fitted dataset.}
\label{fig:LCD}
\end{center}
\end{figure}


In QKD schemes that use polarization encoding, it is necessary to measure the polarization states of the photon stream in at least two different bases.
One commonly used setup employs a rotating wave-plate in front of a polarizing beam splitter.
For an inertial-free basis selection in a small satellite, the rotating wave-plate was replaced by a liquid crystal-based polarization rotator (LCPR).
The optical setup employing the LCPR can be found in \cite{Tang2014}. 


One copy of the LCPR device was exposed to $\gamma$-radiation in a $^{60}$Co chamber at the Centre for Ion Beam Application (CIBA), National University of Singapore.
The device accumulated radiation exposure in steps up to final accumulated value of 1\,krad, which is equivalent to three years of total ionizing dose in the ISS orbit.
At the end of each exposure step, the LCPR is remove from the chamber and recalibrated. The performance of the LCPR in rotating the polarization of a single photon stream is shown in Fig.\,\ref{fig:LCD}(a).

The results presented in Fig.\,\ref{fig:LCD} show that the total transmission and polarization rotation applied by the LCPR for a given voltage are unaffected by the increasing exposed dosage after corrected for the background signals and normalization.
In particular, the polarization extinction ratio was investigated (Fig.\,\ref{fig:LCD}(b)). 
The variation in the observed count rate is attributed to the fluctuation in the single photon source brightness.
We conclude that the LCPR performance is not degraded by the amount of received ionizing radiation.
This also implies that the acceptance level for ionizing damage of the LCPR is way beyond one year of total ionizing dose accumulated in our orbit of interest.
A similar result has been reported for other types of liquid crystal devices \cite{Lane2009}. 
Further investigation on the LCPR performance after exposure to proton beams is being planned.



\section{Conclusion}
\label{sec:conclusion}

In our proposal for space-based QKD, we have identified two types of opto-electronics devices that can be susceptible to space radiation exposure.
The single photon detectors (APDs) were exposed to a proton beam that simulated the displacement damage expected in LEO. 
The APDs demonstrate increasing dark counts with accumulated radiation damage.
However, the measurements on the irradiated APDs suggests that they would operate for the entire life of the satellite (expected to be for one year at an altitude of 400\,km).
The liquid crystal-based polarization rotators (LCPR) were exposed to $\gamma$-radiation and were essentially unaffected by a radiation exposure that is equivalent to three years in orbit.
We conclude that these opto-electronic devices will be able to perform reliably in a space-based QKD experiment in LEO.

\section*{Acknowledgments}
The authors wish to thank Do Thi Xuan Hung for her assistance in the ionizing damage test on the LCPR devices and Prof. T. Osipowicz for access to the $^{60}$Co $\gamma$-chamber. 
\bibliographystyle{tMOP}
\bibliography{ref}

\end{document}